\documentclass[conference]{IEEEtran}
\IEEEoverridecommandlockouts
\usepackage{graphicx}
\usepackage{xcolor}
\usepackage{comment}

\usepackage[hidelinks]{hyperref}

\usepackage[ruled,linesnumbered,vlined]{algorithm2e}

\SetCommentSty{mycommfont}

\usepackage{amssymb}
\usepackage{amsmath}
\usepackage[normalem]{ulem}

\usepackage[caption=false]{subfig}

\usepackage{soul}

\newif\ifcomments
\commentstrue

\newcommand{\IB}[1]{\textcolor{orange}{#1}}
\newcommand{\soutIB}[1]{\IB{\st{#1}}}
\newcommand{\corrIB}[2]{\soutIB{#1} \IB{#2}}

\newcommand \op[1] {\ensuremath{\operatorname{\mathbf{#1}}}}

\newcommand{\hatsub}[1]{\expandafter\hat#1}   

\usepackage[ruled,linesnumbered]{algorithm2e}

\usepackage{amsthm}

\theoremstyle{definition}
\newtheorem{definition}{Definition}[section]
\newtheorem{statement}{Statement}[section]

\newcommand\blfootnote[1]{%
  \begingroup
  \renewcommand\thefootnote{}\footnote{#1}%
  \addtocounter{footnote}{-1}%
  \endgroup
}

\begin{document}
\title{Visual counterexample explanation for model checking with \textsc{Oeritte}}

\author{\IEEEauthorblockN{
Polina Ovsiannikova\IEEEauthorrefmark{1}\IEEEauthorrefmark{2}, Igor Buzhinsky\IEEEauthorrefmark{1}\IEEEauthorrefmark{2}, Antti Pakonen\IEEEauthorrefmark{3}, and Valeriy Vyatkin\IEEEauthorrefmark{1}\IEEEauthorrefmark{2}\IEEEauthorrefmark{4} }
\IEEEauthorblockA{\IEEEauthorrefmark{1}Computer Technologies Laboratory, ITMO University,
Saint Petersburg, Russia}
\IEEEauthorblockA{\IEEEauthorrefmark{2}Department of Electrical Engineering and Automation,
Aalto University, Espoo, Finland}
\IEEEauthorblockA{\IEEEauthorrefmark{4}Department of Computer Science, Computer and Space Engineering,
Lulea Tekniska Universitet, Sweden}
\IEEEauthorblockA{\IEEEauthorrefmark{3}VTT Technical Research Centre of Finland Ltd.,
Espoo, Finland \\ Email: polina.ovsiannikova@aalto.fi, igor.buzhinskii@aalto.fi, antti.pakonen@vtt.fi, valeriy.vyatkin@aalto.fi}
}

%


\maketitle              
\blfootnote{The 25\textsuperscript{th} International Conference on Engineering of Complex Computer Systems (ICECCS 2020). © IEEE. Personal use of this material is permitted. However, permission to reprint/republish this material for advertising or promotional purposes or for creating new collective works for resale or redistribution to servers or lists, or to reuse any copyrighted component of this work in other works must be obtained from the IEEE.}
\begin{abstract}
Despite being one of the most reliable approaches for ensuring system correctness, model checking requires auxiliary tools to fully avail. In this work, we tackle the issue of its results being hard to interpret and present \textsc{Oeritte}, a tool for automatic visual counterexample explanation for function block diagrams. To learn what went wrong, the user can inspect a parse tree of the violated LTL formula and a table view of a counterexample, where important variables are highlighted. Then, on the function block diagram of the system under verification, they can receive a visualization of causality relationships between the calculated values of interest and intermediate results or inputs of the function block diagram.
Thus, \textsc{Oeritte} serves to decrease formal model and specification debugging efforts along with making model checking more utilizable for complex industrial systems.




\end{abstract}


\section{Introduction}
One of the biggest advantages of \emph{model checking}~\cite{clarke1999} is a possibility to ensure that the specification is satisfied in every state of the given formal model. This fact is especially useful when it comes to verification of systems on chips~\cite{kaivola2009intel}, the Internet of things installations~\cite{nakahori2017iot} and other safety-critical systems. Specifically, in this work, we will emphasize its application for industrial size instrumentation and control (I\&C) systems. Model checking has been successfully used for verifying nuclear power plant (NPP) I\&C system design in Finland~\cite{pakonen2017npic}, Korea~\cite{Jee2010}, Hungary~\cite{nemeth2009}, and at the European Organization for Nuclear Research (CERN)~\cite{adiego2015}. It was also used to check safety of aircraft flight control~\cite{gelmanAirbus2013} systems.
The distinctive features of I\&C systems with regard to model checking are their complexity, modularity and the need to comply with safety requirements, which means that every failure result should be thoroughly analyzed and fixed.

However, in its initial version, model checking is not user-oriented and its application requires additional knowledge about a system as a whole, as well as time and efforts aimed to localize an error in the model of the system. The tool \textsc{Oeritte}, presented in this paper, does one step closer to the user-friendly model checking and focuses on explanation of negative verification results.


In this work, we consider I\&C modular models specified as \emph{function block diagrams (FBD)}.\footnote{The presented contributions are not limited to FBDs as specified in the IEC 61131-3 standard~\cite{IEC61131}. Instead, we use this term in a more general sense as described in this paragraph and later.}
An FBD is an arrangement of interconnected entities called \emph{function blocks}.
Each function block can be assumed to be a deterministic Mealy finite-state machine~\cite{lee2016introduction}.
We use the synchronous FBD execution semantics: on each discrete step, each block executes once, connections link the outputs of blocks to inputs of other function blocks already on the current time step, and feedback loops are manually broken with delay function blocks to prevent the infinitely fast flow of information.
Practically, the FBDs we work with can be constructed with graphical tools such as MODCHK~\cite{pakonen2017esrel} or Simulink Design Verifier~\cite{simulink}, but they can also be specified in the languages of model checkers, such as NuSMV~\cite{nusmv}.

The set of properties to be checked for the system, or its specification, is assumed to be formulated in \emph{linear temporal logic} (LTL)~\cite{clarke1999}.
This logic is a generalization of the propositional Boolean logic over state sequences, where a \emph{state} is an assignment of values to the variables of the checked formal model.
In LTL, the following \emph{temporal operators} are used (below, $\varphi_1$ and $\varphi_2$ are LTL formulas): 
\begin{itemize}
    \item $ \op G \varphi_1$ (``globally''):  $\varphi_1$ must be true on the entire trace of the model;
    \item $ \op F \varphi_1$ (``finally''):  $\varphi_1$ must hold eventually;
    \item $\varphi_1 \op U \varphi_2$ (``until''): $\varphi_1$ must be true until $\varphi_2$ is true, and the latter is required to eventually happen; 
    \item $ \op X \varphi_1$ (``next''): $\varphi_1$ must be true for the next state.
\end{itemize}

A finite or infinite state sequence is \emph{valid} if it starts in one of the model's initial states and every pair of adjacent states belongs to the model's transition relation.
An \emph{LTL formula} (also called an LTL property) is assumed to be satisfied for the model when it is satisfied for all its valid state sequences.
\emph{Model checking} of an LTL formula constitutes finding whether the formula is satisfied for the model and, if it is not, finding a \emph{counterexample} that demonstrates its violation.

Now, a \emph{counterexample}~(or a failure trace) for a property $\varphi$ in a given formal model is a valid state sequence where $\varphi$ is not satisfied. In model checking, it is sufficient to consider finite counterexamples and the ones that can be represented as a prefix combined with a loop (so-called lasso-shaped counterexamples). 

Counterexamples are usually represented by a table of values and for industrial systems are often incomprehensible~\cite{jee2010fbdverifier} without auxiliary tools. 
Therefore, we outline three challenges that analysts usually face when trying to utilize a failure trace in the verification process:
\begin{enumerate}
    \item For industrial I\&C system models, counterexample traces could be quite long (tens of steps)~\cite{pakonen2018} due to the delay elements in the logics, necessary for implementing complex control sequences. In the meantime, each of steps may include dozens of variables where inputs and outputs are usually mixed. Hence, without having a counterexample visualized, it can be difficult to pinpoint on which time step the property fails. 
    \item Specification failures are neither obvious to localize, i.e., a property can be of high complexity, requiring pen and paper to evaluate it and identify the branch that has caused the problem.
    \item When it comes to faults of the system itself, the analyst has to account for every module to figure out the source of the problem, which requires the analyst to have strong expertise in the formal language used for the system's implementation.
\end{enumerate}

\textsc{Oeritte} assists the analyst in three ways. First, it visualizes FBD data flow throughout the counterexample. Second, it uses the algorithm from~\cite{pakonen2018} to explain a violated LTL property. Third, it implements a novel algorithm that explains individual assignments in terms of highlighted assignments and connections between them in a given FBD.

The rest of the paper is structured as follows. Section~\ref{sec:fbd} describes the internal representation of an FBD in \textsc{Oeritte}. The problem of counterexample explanation and algorithm that solves it are given in Sections~\ref{sec:problem} and~\ref{sec:alg}. Section~\ref{sec:tool} overviews the tool that implements the proposed approach, and Section~\ref{sec:case_study} evaluates the approach experimentally. Related research is reviewed and compared with the proposed approach in Section~\ref{sec:rel}. Section~\ref{sec:conclusion} concludes the paper.

\section{System representation as FBD}
\label{sec:fbd}


In this work, function blocks of an FBD that represents a system under verification can be of two types: basic and complex. A \emph{basic block} is a block corresponding to some atomic operator or a simple function listed in Table~\ref{tab:blocks}. \emph{Complex blocks} operate with Booleans and integers and can be decomposed into combinations of interconnected function blocks. An FBD, which is a system under verification, is a complex block of the highest hierarchy level, i.e., it is not a part of any other complex block. When we say \emph{blocks}, we consider both types of blocks~-- basic and complex.

Blocks have input and output interfaces that are comprised of named \emph{gates}.
A gate that belongs to the interface of a complex block can have single incoming and multiple outgoing connections. On the other hand, for basic blocks, input gates cannot have outgoing connections, and output gates cannot have incoming ones.
We say that blocks \emph{A} and \emph{B} are connected if there is a gate in \emph{A} that has an outgoing connection \emph{c} and there is a gate in \emph{B} that has connection \emph{c} as an incoming one~(or vice versa).

\begin{table}[b]
\centering
\caption{Basic blocks used for the module representation. }\label{tab:blocks}
\begin{tabular}{p{3cm}p{5cm}}
\hline
Logical &  $\land$, $\lor$, $\Leftrightarrow$ \\
\hline
Arithmetical & $-$, $+$, $\times$, $\div$ \\
\hline
Relation &  $>$, $<$, $\leq$, $\geq$, $=$ \\
\hline
Other & \texttt{DELAY}, \texttt{CHOICE}, \texttt{COUNT}, \texttt{ASSIGN} \\
\hline
\end{tabular}
\end{table}

Now let us consider the types of basic blocks listed in Table~\ref{tab:blocks}. The meaning of the blocks in the first three groups is quite straightforward and their interfaces are comprised of two input gates and one output gate.
Blocks from the group ``Other'' require some elaboration:
\begin{itemize}
    \item {With the \texttt{DELAY} basic block, one can construct modules with memory, i.e., if calculations at the next time step require the value that was inferred at one of the previous ones, it is possible to delay it by one cycle and have it as an input in the same complex block in the future. This block has two inputs~-- one for the variable that should be delayed and one for its default value at the first step. Delay by $N>1$ cycles can be implemented as a chain of $N$ \texttt{DELAY} blocks.}
    \item The \texttt{CHOICE} basic block imitates a chain of conditional assignments, following the semantics of the \texttt{case} NuSMV operator. It ends with the ``else'' branch defining the value in case none of the conditions are satisfied. The inputs of this block are conditions and the outputs that correspond to them. The output of \texttt{CHOICE} is the value of the output for the first satisfied condition.
    \item The output of the \texttt{COUNT} block is the number of input signals that are \texttt{TRUE} at the current step. This block requires that its inputs are Boolean.
    \item The \texttt{ASSIGN} block implements the identity function: the output is the same as the input.
\end{itemize}

\begin{figure}[t]
    \centering
    \includegraphics[width=0.5\textwidth]{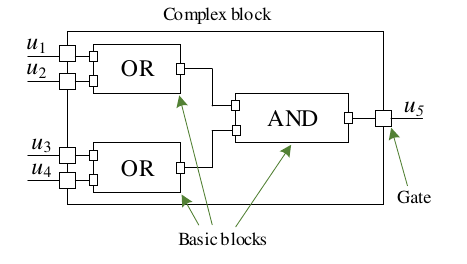}
    \caption{A complex block that encodes function $u_5 = (u_1 \lor u_2) \land (u_3 \lor u_4)$.}
    \label{fig:block:insides}
\end{figure}

Connections between basic blocks can be usual and inverted, i.e., the accepting block receives the inverted signal from the output of the producing block. 
Hence, due to having $\land$, $\lor$ blocks and inversion, the chosen set of basic blocks is sufficient to express any Boolean function. 
An example of a complex block with its internal structure is provided in Fig.~\ref{fig:block:insides}.

\section{Counterexample explanation problem}
\label{sec:problem}
Informally, we aim to explain the false outcome of an LTL formula $\varphi$ on counterexample $X$ of length $l$ to a given FBD $D$ with its set of variables $U = \{u_1,...,u_n\}$ using both the values of state variables of the counterexample and the blocks in $D$.
Formally:

\begin{definition}[Assignment]
An \emph{assignment} $a$ is a tuple $(u, v_{u,j}, j)$, where $v_{u,j}$ is the value of variable $u$ at discrete time step $j$. By $v(a)$ we denote the value of this assignment and by $s(a)$ its step. If $u \in U$ is a variable of $D$ then there exists index $i \in [1,n]$ for $u$, and, to simplify the text, assignments for $u$ instead of $a_{u_i,j}$ are denoted as $a_{i,j}$.
\end{definition}

\begin{definition}[Counterexample]
A \emph{counterexample} $X$ of length $l$ is a set of assignments of the variables from $U$ for each time step: $X =\{(u_i,v_{i,j},j) \:|\: i \in [1,n], j \in [1,l]\}$. 
\end{definition}

FBD $D$ consists of complex blocks which, in the end, are decomposed into nets of basic blocks. 
We view basic blocks as sets of symbolic constraints on their input and output variables. These constraints are specified for each instance of each basic block.

\begin{definition}[Basic block constraints]
\label{def:b_constr}
Every basic block $B$ with $k$ input variables and one output variable from Table~\ref{tab:blocks} except for \texttt{DELAY} is uniquely defined by its set of constraints $C_B = \{v_{1,s} = f_B(v_{2,s},...,v_{k,s}) \:|\: s \in [1,l]\}$, where $s$ is a step of $X$, $f_B$ is determined by the type of each basic block, and $u_2,...,u_k$ and $u_1$ are $k-1$ input and one output variables of a particular instance of $B$. \texttt{DELAY} corresponds to the following set of constraints: $C_B = \{v_{1,1}=v_{2,1}\} \cup \{v_{1,s} = v_{3,s-1} \:|\: s \in [2,l]\}$, where $v_{2,1}$ is a default value for the first counterexample step, and $u_1,..,u_3$ are variables 
of a particular instance of \texttt{DELAY}.
\end{definition} 

For example, the set of constraints for the \texttt{AND} block is $C_\land = \{v_{1,s} = v_{2,s} \land v_{3,s} \:|\: s \in [1,l]\}$. Each constraint in such a set encodes a rule of how the block functions at a particular counterexample step and, therefore, the number of elements in the set equals to the length of a counterexample. 
Connections also correspond to similar constraints, thus making the constraints of all blocks linked to each other:

\begin{definition}[Connection constraints]
\label{def:con_constr}
Assume that output variable of some block $u_{1}$ is connected to input variable $u_{2}$ of another block. Then a set of constraints for such connection $c$ is $C_c = \{v_{1,s}=v_{2,s} \:|\: s \in [1,l]\}$.
\end{definition}

\begin{definition}[Complex block constraints]
\label{def:cb_constr}
Let $B$ be a complex block with its set of internal blocks $M$ and set of internal connections $\Sigma$. Then, a set of constraints for $B$ is $C_B = \{C_{m} \:|\: m \in M\} \cup \{C_{\sigma} \:|\: \sigma \in \Sigma \}$.
\end{definition}

\begin{figure}[b]
    \centering
    \includegraphics[width=0.5\textwidth]{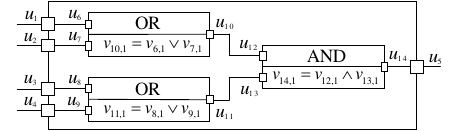}
    \caption{Complex block $B$ from Fig.~\ref{fig:block:insides} with constrains of its internal blocks for the first counterexample step defined. The full set of constraints for $B$ for the first counterexample step is represented by the union of constraints for the depicted basic blocks and the set of connection constraints $C_c=\{v_{6,1}=v_{1,1},v_{7,1}=v_{2,1}, v_{8,1}=v_{3,1},v_{9,1}=v_{4,1},v_{12,1}=v_{10,1},v_{13,1}=v_{11,1},v_{5,1}=v_{14,1}\}$, where $v_{i,j}$ is a value of variable $u_i$ at counterexample step $j$, $i \in [1,|U|], j \in [1,l]$. }
    \label{fig:block:constr}
\end{figure}

As an FBD is a complex block, the set of constraints $C_D$ is defined for $D$ as well.
Definitions~\ref{def:b_constr},~\ref{def:con_constr} and~\ref{def:cb_constr} are illustrated in Fig.~\ref{fig:block:constr} for a counterexample of length 1.
Fig.~\ref{fig:block:constr} shows how the set of constraints for the complex block from Fig.~\ref{fig:block:insides} can be defined. 

Due to the possibility of explaining the outcome of $\varphi$ through assignments of variables that is present in it~\cite{beer2012,pakonen2018}, we can decompose the process of explaining the outcome of $\varphi$ to the one of explaining a number of individual assignments in the counterexample.
Below, we focus on explaining a single assignment, called an \emph{explanation target}.
The explanation target can be represented by an input or output assignment of any block structure: an FBD, a complex block, or a basic block.
Initially, explanation targets come from applying the approach~\cite{pakonen2018}, but we also allow the situation where the user selects a custom explanation target to focus on a particular part of $D$, thus allowing more flexibility in explanation.



\begin{definition}[Cause]
A set of assignments $C \subseteq X$ is a cause of a target $t$ if there exists such sequence of sets of assignments from $X$, $Y_0,...,Y_m : C = Y_0, t \in Y_m$, where each $Y_{k+1},k \in [0, m-1]$ extends $Y_k$ with a single assignment $a_{i,j}' \in X$, there exists constraint $c^* \in C_D$ such that the formula
\begin{equation}
c^* \land \left(\bigwedge\limits_{a_{i,j} \in Y_k} (v_{i,j} = v(a_{i, j}))\right) \to \left(v_{a_{i,j}'}=v(a_{i,j}')\right) \label{eq:cause}
\end{equation} 
is valid, and $a'_{i, j}$ refers to the output variable of the basic block or connection to which $c^*$ corresponds.
\label{def:cause}
\end{definition}

Intuitively, in every set $Y_{k}$ from the definition above there exists a cause of the new assignment that is added to $Y_k$ to obtain $Y_{k+1}$ and at some extension step $q < m$, $t$ should be added to get $Y_{q+1}$. 

This definition can also be explained in terms of logical inference. Suppose that each statement is an assignment. Then the definition says that it is possible to infer $t$ given a set of statements $C$ if the allowed rules are limited to using input-output dependencies of each individual basic block or connection in the direction of the information flow.

\begin{definition}[Inclusion-minimal cause]
$C \subseteq X$ is an \emph{inclusion-minimal cause} of $t$ if $C$ is a cause of $t$ and there is no $C' \subset C$ that is a cause of $t$.
\end{definition}

Having these definitions, we say that to \emph{explain the target}~(or to find a cause of the target) means to find the union of its inclusion-minimal causes.

Here, it is worth mentioning the commonly accepted notion of causality from~\cite{halpern2015modification}. This work defines \emph{actual} but-for causes of $\varphi$ under some contingency in the model represented by structural equations. Meanwhile, our causes are not necessarily counterfactual. In other words, we deal with general causation, considering a set of assignments as a cause of $t$ if it is sufficient to infer $t$ in the given FBD.

As an example, consider the basic block \texttt{AND} from Fig.~\ref{fig:block:constr} and a counterexample of length 1. Assume that the explanation target is $t=(u_{14},0,1)$, and $v_{12,1}=1$, $v_{13,1}=0$~(we denote logical values \texttt{TRUE} and \texttt{FALSE} as 1 and 0 respectively). To find out if any of input variables $U=\{u_{12},u_{13}\}$ of \texttt{AND} are included in a cause of $t$, we, first, substitute $c^*$ in~(\ref{eq:cause}) with $v_{14,1}=v_{12,1} \land v_{13,1}$. Then, as soon as $U$ and $t$ belong to the same basic block without delay, the only one constraint is required to infer the cause, hence, the length of the sequence of sets from Definition~\ref{def:cause} is two, where the first one is a cause.
Now, we rewrite~(\ref{eq:cause}) as
\begin{equation}
\begin{split}
    \left(v_{14,1}=v_{12,1} \land v_{13,1}\right) \land \left(\bigwedge\limits_{a_{i,1} \in C} (v_{i,1} = v(a_{i, 1}))\right) \\ \to (v_{14,1} = 0),\label{eq:cause_ex}
\end{split}    
\end{equation}
where $i$ in the middle part is an index of the variable from $U$.

Having~(\ref{eq:cause_ex}), the next step is to pick such assignments for $C$ so that the relation~(\ref{eq:cause_ex}) is valid and $C$ is inclusion-minimal. In this example, there exists one such set of assignments $C = \{(u_{13},0,1)\}$.

With the set of assignments $U$ that can be potentially but not necessary added to $Y_0$ from Definition~\ref{def:cause} in~(\ref{eq:cause}), it is possible to set an \emph{explanation scope}. 
In the previous example, the scope was defined by the input assignments of \texttt{AND} at step 1.
Alternatively, if we explain $t$ using input assignments of both \texttt{OR} blocks at the same step, constraints for all basic blocks shown in Fig.~\ref{fig:block:constr} and two constraints for the connections $\{v_{10,1}=v_{12,1},v_{11,1}=v_{13,1}\}$ will be used in the extension procedure. Assume $v_{8,1}=0$ and $v_{9,1}=0$. Then the chosen scope produces the following inclusion-minimal cause: $C=\{(u_8,0,1),(u_9,0,1)\}$. 

\section{Assignment explanation algorithm}
\label{sec:alg}



The problem stated in Section~\ref{sec:problem} assumes that among all system assignments an inclusion-minimal cause of an explanation target should be found. To do this, first, we define a global explanation scope as the union of all input assignments of the FBD that the explanation target belongs to and assignments inside the FBD that have names of the gates which do not have incoming connections. Next, for any complex block it is a dubious help to see how, e.g., its output depends on its inputs, the analyst usually wants to know why such dependency takes place. Hence, in the explanation result we also include inclusion-minimal causes for every nested explanation scope if they exist for such scope. Thirdly, sometimes (for complex blocks) there can be more than one inclusion-minimal cause and it is the user who chooses the one of their interest, thus, we need to discover the union of all such causes. 


\begin{algorithm}[t]
\SetNoFillComment
\SetKwIF{If}{ElseIf}{Else}{if}{then}{else if}{else}{}
 \KwData{FBD $D$, counterexample $X$, explanation target $t \in X$}
 \KwResult{set $C$~-- the union of all inclusion-minimal causes of $t$ in $D$}
 
 \If{\upshape $t$ corresponds to an input variable of $D$ or a constant block input}{
    \Return{$\{t\}$} \tcp{this is a terminating cause}
  }
  \ElseIf{\upshape $t$ is an input variable of a basic block in $D$}{
    \tcc{follow the connection and add it to the tree}
    $t' \leftarrow$ the assignment of the output variable at the opposite end of the connection where $t$ is located\\
    $C_0 \leftarrow \mathtt{explain}(D, X, t')$\\
    \Return{$C_0 \cup \{t\}$}\\
  }
  \Else(\tcp*[h]{$t$ is an output variable of some basic block}){ 
    $t_1, ..., t_m \leftarrow$ causes found for the current basic block type according to Table~\ref{tab:exprules}\\
    $C \leftarrow \{t\}$\\
    \tcc{recursively explain the assignments of the local cause}
    \For{$i=1$ \KwTo $m$}{
      $C_0 \leftarrow \mathtt{explain}(D, X, t_i)$\\
      $C \leftarrow C \cup C_0$\\
    }
    \Return{$C$}\\
  }
\caption{Assignment explanation algorithm \texttt{explain}.}
\label{alg:explain}
\end{algorithm}

\subsection{FBD preprocessing}

While complex blocks of the lowest hierarchy level in the FBD can be decomposed into basic blocks, this is not done in the original FBD. Thus, the values of the internal variables of such complex blocks are missing in a counterexample, although they are required for the explanation procedure. To obtain an extended counterexample, before running the algorithm for target $t$ on FBD $D$, decomposition is performed automatically, and the full set of constraints for each of mentioned complex blocks is added to the full constraint set of the diagram $C_D$ and the values of new variables are calculated for each counterexample step $s\in[1,s(t)]$. 

This stage also provides a way to ensure that the complex block is parsed correctly, as otherwise, after execution, its output variable values could differ from the ones stated in the counterexample.

\subsection{Recursive explanation}

\begin{figure*}[t]
\centerline{\includegraphics[width=\textwidth]{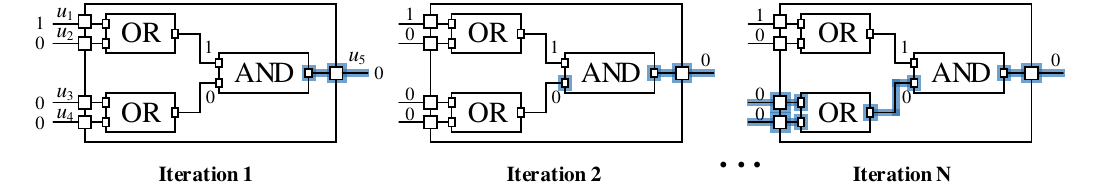}}
\caption{Illustration of the assignment explanation process. Digits above the connections show values of the transmitted signals.}
\label{fig:expsteps}
\end{figure*}

The algorithm is provided in Alg.~\ref{alg:explain} and is illustrated in Fig.~\ref{fig:expsteps}, where the problem is to explain why output variable $u_5$ of the complex block is \texttt{FALSE} at counterexample step $s$.

Recalling that an FBD itself is a complex block of complex blocks, to explain its output assignment, we need to find the output gate connected to the gate of the output of interest in the nested complex or basic block~(Fig.~\ref{fig:expsteps}, iteration~1). Then, if the found gate belongs to a complex block, the output assignment of such a block is explained through the underlying net of blocks, whereas to explain an output of a basic block, rules from Table~\ref{tab:exprules} are utilized. 
As a result, we have a set of input assignments that are sufficient to make explained basic block output have its particular assignment~-- an inclusion-minimal cause~(Fig.~\ref{fig:expsteps}, iteration~2). 
If the obtained inputs have incoming connections, we continue the explanation procedure recursively in the same way; intermediate results from each step are added to the overall result set.
After the algorithm terminates, the result composed of all inclusion-minimal causes in global and all the nested explanation scopes is obtained~(Fig.~\ref{fig:expsteps}, iteration~$N$) and its graphical visualization described in Section~\ref{sec:expl_mode}.

The time and memory complexity of the algorithm is $O(n \cdot s(t))$, where $n$ is the number of variables in the FBD~(including ones that belong to internal basic blocks).
These estimates can be achieved if the result of each call of \texttt{explain} is memorized and not recomputed. 

\begin{figure}[b]
\centerline{\includegraphics[width=0.5\textwidth]{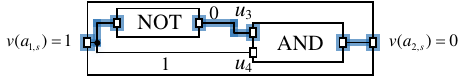}}
\caption{Illustration of a non-intuitive algorithm result at step $s$ due to gates $u_3$ and $u_4$ having a common ancestor $u_1$. The path of an explanation process for $a_{2,s}$ is highlighted with bold blue. Here block \texttt{NOT} inverts the signal from gate $u_1$, thus, block \texttt{AND} computes the expression $v(a_{2,s}) = v(a_{1,s})  \land \lnot v(a_{1,s})$, which is always false. The algorithm will result in the path in bold blue and $\{(u_1,1,s)\}$ will be the inclusion-minimal cause of $u_2$. 
} 
\label{fig:falsepositive}
\end{figure}

\begin{statement}
Alg.~\ref{alg:explain} finds the union of all inclusion-minimal causes of $t$.
\label{stm:correctness}
\end{statement}

For brevity, we leave this statement without a formal proof and only give an intuitive explanation why it is satisfied.
The algorithm performs a backward (in terms of the information flow in the FBD) cone-of-influence analysis, seeking for all assignments that could possibly be the cause of $t$ according to Definition~\ref{def:cause}.
Note that this definition requires that any cause must be sufficient to reach the target by inferring new assignments only in the direction of the information flow, which means that a search against this flow could reach all these causes.
Moreover, the rules in Table~\ref{tab:exprules} were specifically chosen to return the union of inclusion-minimal causes for an output of an FBD composed of an isolated basic block.
For arbitrary FBD, Statement~\ref{stm:correctness} could be proven by induction.

\begin{table}[b]
\centering
\caption{Basic block explanation rules of finding local inclusion-minimal causes. Assume that the request is the explanation target represented by the tuple $(u,v,s)$, where $v$ is the value of $u$ at step $s$, and a set of assignments representing a cause is returned.}\label{tab:exprules}
\begin{tabular}{p{2cm}p{6cm}}
\hline
Block & Rule\\\hline
Logical \texttt{AND} & If $v$ is \texttt{TRUE}, then return all the block input assignments for step $s$, else return only input assignments that are \texttt{FALSE} at $s$. \\
\hline
Logical \texttt{OR} & If $v$ is \texttt{FALSE}, then return all the block input assignments for $s$, else return only inputs that are \texttt{TRUE} at $s$. \\
\hline
\texttt{CHOICE} & Return all the condition assignments prior to and including the one that is satisfied at $s$ and its output assignment. \\
\hline
\texttt{DELAY} & Return input assignment from step $s-1$. \\
\hline
Others & Return all the input assignments for step $s$. \\
\hline
\end{tabular}
\end{table}

\subsection{Discussion}

The algorithm finds the union of all inclusion-minimal causes for the target within the global and all the nested explanation scopes. Whereas our definition does not consider the system as a whole at every extension step, there may exist such combinations of constraints that generally restrict combinations of values of input assignments of basic blocks. For instance, consider Fig.~\ref{fig:falsepositive}, where signals merge in a common ancestor if traversing backwards from $a_{2,s}$.

Another case is an explanation result of basic block \texttt{COUNT} that includes all the input assignments and is inclusion-minimal. It follows that further explanation results will include causes for all of the input assignments of \texttt{COUNT} which may be irrelevant with regard to the system structure. Assume that the output of this block equals 4 and is connected to an input of ``$\le$'' block, where it is compared to 6. Having this context in mind, we know that inputs of our interest are \texttt{FALSE} because two more \texttt{TRUE} signals are required to change the ``$\le$'' output.
This example shows that without the knowledge of how the result of this block is used further, it is hard to say should we consider its \texttt{TRUE} or \texttt{FALSE} signals in the explanation. Nevertheless, assignments that are not included into the result of the algorithm are insufficient in the target explanation, therefore, the user still gets information that significantly reduces an FBD area being analyzed.


\begin{figure*}[t]
\centerline{\includegraphics[width=1.0\textwidth]{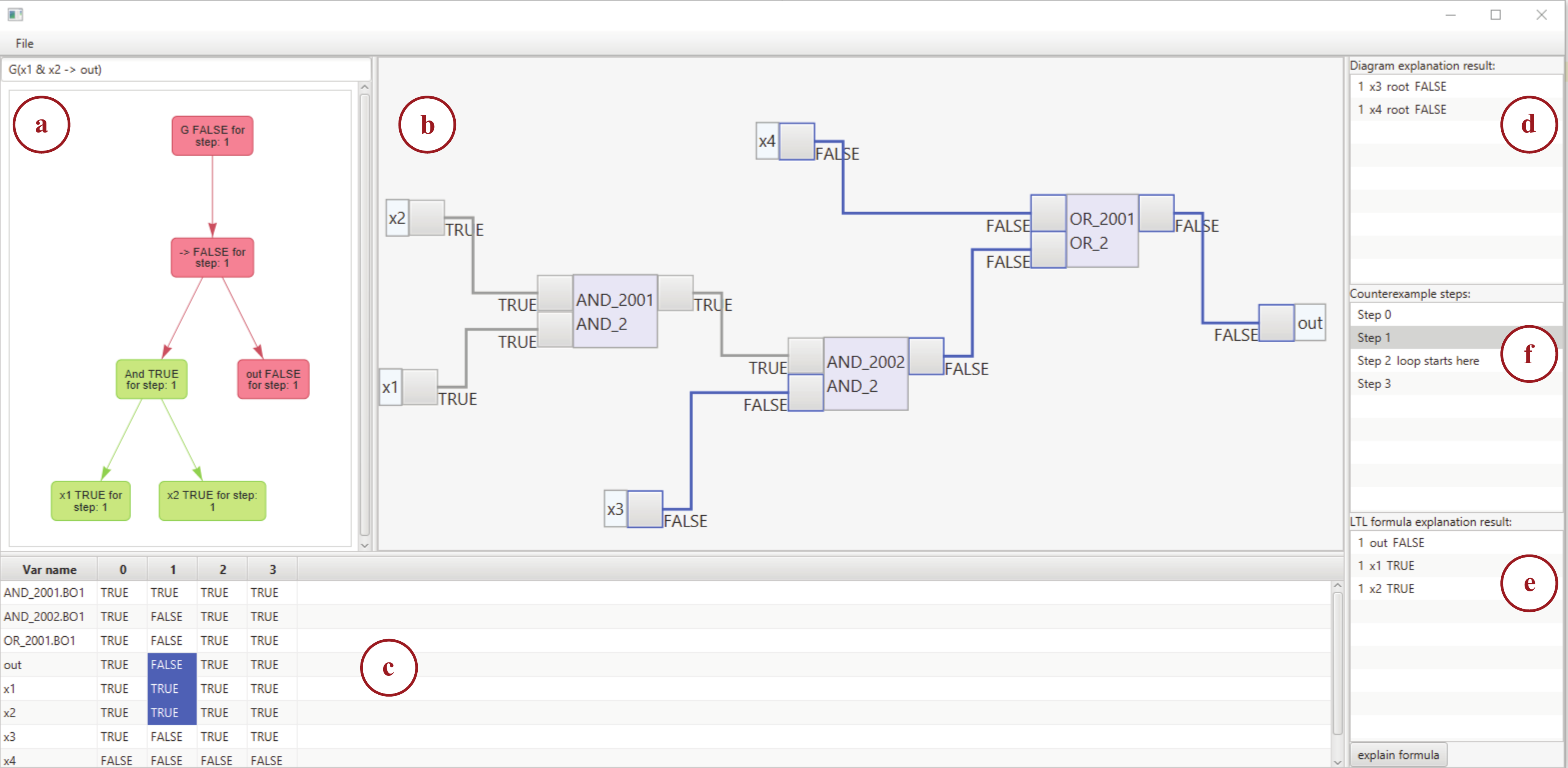}}
\caption{Main view of the tool in the explanation mode.}
\label{fig:main_view}
\end{figure*}

\section{Implementation}

The implementation of the algorithm described in Section~\ref{sec:alg} was incorporated into the tool \textsc{Oeritte} with the user interface developed to aid the analyst in the debugging process.

\label{sec:tool}
\subsection{Input data}
\label{sec:input_data}
The tool accepts a NuSMV model, an LTL formula and a counterexample for the provided formula on the provided model as input.
A restricted, but, nonetheless, already usable according to our practical experience, subset of NuSMV and LTL is supported.
Below are the main limitations:
\begin{itemize}
\item The \texttt{main} module of the NuSMV model is restricted to declarations of input variables and nested modules.
\item In other modules, each internal variable must be declared with \texttt{init} and \texttt{next} operators.
These assignments must be deterministic (set notation $\texttt{\{...\}}$ is disallowed).
\texttt{INIT} and \texttt{TRANS} declarations are not allowed.
\item \texttt{DEFINE} declarations are not allowed to use the \texttt{next} operator. 
\item Only Boolean and integer scalar types are supported.
\item Inputs of the NuSMV modules should be annotated with their types in form ``\texttt{varName}~:~\texttt{type}'', where \texttt{type} is \texttt{boolean} for Boolean and any integer interval in form \texttt{start..end} for integer, e.g., \texttt{0..100}.
\item In LTL formulas, bounded operators (e.g., \texttt{G[0, 3]}) and past time operators (e.g., \texttt{H}) are not supported.
\end{itemize}

\subsection{Encoding NuSMV modules as complex blocks}

The aforementioned determinism assumption is required to represent NuSMV modules as complex blocks since our basic blocks are purely deterministic.
The input variables of the complex block correspond to input variables of the module, and the output variables correspond to its internal variables and \texttt{DEFINE} declarations (the absence of \texttt{next} operators inside them allows treating these declarations as if they were internal variables).
Logical and arithmetic NuSMV operations are directly transformed to basic blocks listed in
Table~\ref{tab:blocks}.
To handle delays introduced with the \texttt{next} operator, we create a delayed version of each input variable by passing it through a \texttt{DELAY} block.
Each output variable is then wired to a \texttt{CHOICE}, which, depending on whether this is the first cycle, outputs the \texttt{init} or the \texttt{next} expression for this variable: \texttt{init} expressions always use undelayed variables, while \texttt{next} expressions may use both undelayed and delayed ones.

\subsection{Main window overview}
The main view of \textsc{Oeritte}\footnote{\url{https://github.com/ShakeAnApple/cxbacktracker/}} is presented in Fig.~\ref{fig:main_view}.  
Here two areas are used to depict an LTL formula tree~(Fig.~\ref{fig:main_view}a) and the visualized version of an FBD~(Fig.~\ref{fig:main_view}b) (hereinafter, the \emph{diagram}). The table below them~(Fig.~\ref{fig:main_view}c) shows the values of all the system variables for every counterexample step, and the discovered failure causes for the diagram and the LTL formula can be found in lists~(Fig.~\ref{fig:main_view}d) and~(Fig.~\ref{fig:main_view}e) correspondingly. 
Panel~(Fig.~\ref{fig:main_view}f), containing the trace steps sequence, is clickable and utilized for navigation: once an element is activated, both the diagram and the formula tree are evaluated according to the chosen step, so that all variables in both diagrams are assigned with values defined by the counterexample step, hence, all the nodes in the LTL formula tree are calculated and all the system modules are executed. 

\textsc{Oeritte} has two groups of features: LTL formula explanation and diagram interpretation. 
For the first, we implemented the cause identification algorithm from~\cite{pakonen2018}. 
The LTL formula parse tree~(see Fig.~\ref{fig:main_view}a) is located on the leftmost panel of the main view. 
Depending on the calculation result of the branch, the nodes of the tree are colored in red, green and grey for \texttt{FALSE}, \texttt{TRUE} and an arithmetic result respectively. 
The ``explain formula'' button forces the causes  of the formula value at the current step to appear in the list on the right and be colored in the value table.
By default, the formula is explained for the first step (step 0 in the tool) with the first diagram evaluation.

\subsection{Complex block}
In Fig.~\ref{fig:block:visual}, an example of such visual representation of a NuSMV module is given. 
This block is a part of a diagram in the explanation mode. 
Each block has a \emph{name} and a \emph{type}~(Fig.~\ref{fig:block:visual}a). 
Two sets of pins on the left~(Fig.~\ref{fig:block:visual}b) and right~(Fig.~\ref{fig:block:visual}c) sides are the block's inputs and outputs that together form its interface. 
If the diagram is not in the explanation mode, hovering the pin will trigger a tooltip with the variable name appearing. 
Blocks with the single input or output pin on the left and right sides of the diagram represent a \emph{system interface}~-- input and output variables of the whole system. 
Lines connecting module inputs and outputs correspond to connections between the blocks. 
Output variable values are shown near the output connecting points of these lines, whereas input variable values of the blocks are placed near the input connecting points.
Input negation is represented as a circle instead of a square~(Fig.~\ref{fig:block:visual}d).

\subsection{Explanation mode}
\label{sec:expl_mode}
\begin{figure}
    \centering
    \includegraphics[width=0.4\textwidth]{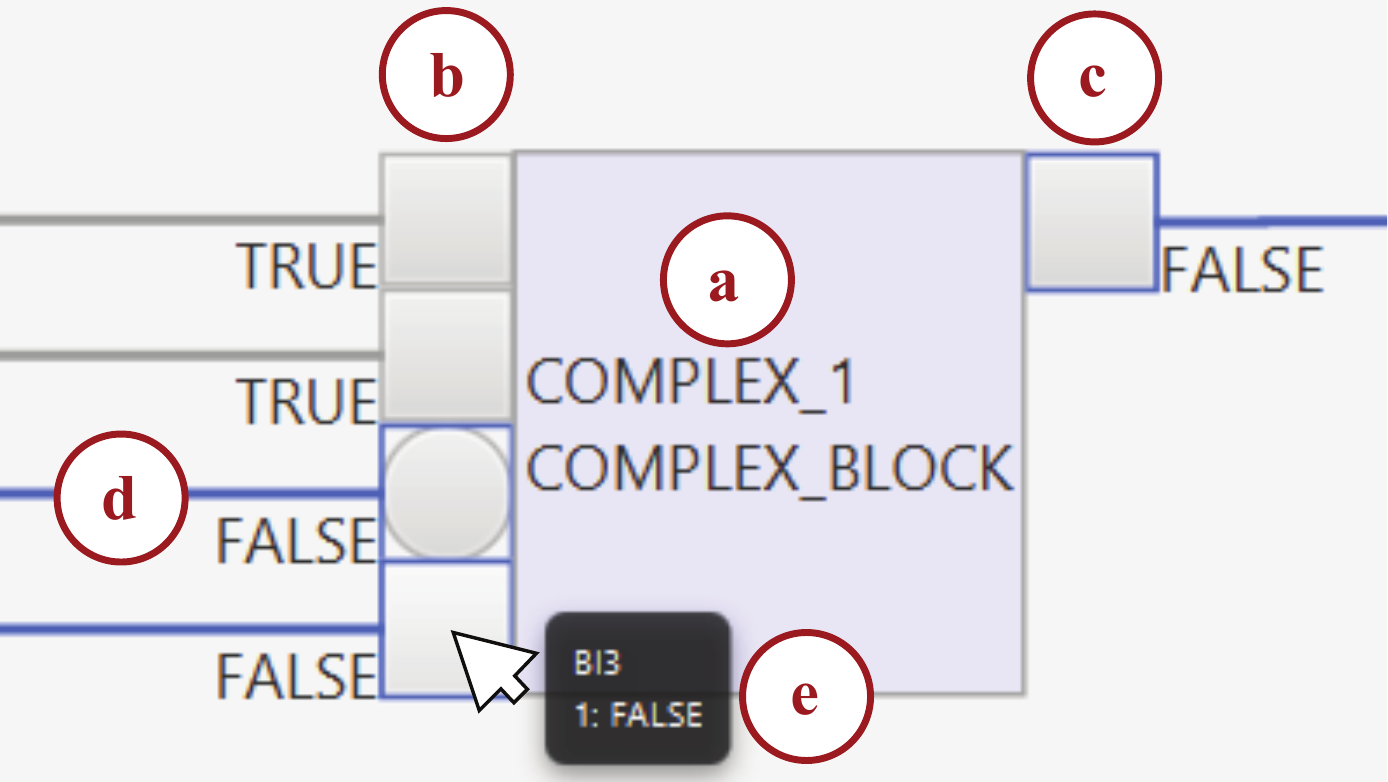}
    \caption{Visual representation of a complex block in the explanation mode.}
    \label{fig:block:visual}
    \vspace{-0.4cm}
\end{figure}

Visualization of the assignment explanation algorithm result is shown in Fig.~\ref{fig:main_view} and is done as follows. 
To switch to the explanation mode, one needs to click the desired pin at the desired step or explanation target on the diagram view, which triggers the aforementioned backward explanation process that results in a set of assignments belonging to a calculated cause. 
To display such a set on the diagram view, we hide the time dimension and highlight edges that connect output and input pins from the common set of causes with blue.  At  the  same  time,  if  some  variable  is  a  cause  at  several time steps, the pin representing this variable is provided with a tooltip where all its values included in the cause are displayed in form  ``\texttt{stepNumber:value}'' (Fig.~\ref{fig:block:visual}e). 
Together with graphical visualization, list~(Fig.~\ref{fig:main_view}c) shows \emph{terminating assignments}, i.e., assignments, whose gates do not have incoming connections and belong to the input interface of the system. They are displayed in form ``\texttt{stepNumber} \texttt{varName} \texttt{blockName} \texttt{value}''. This is the output of our interest, which 
includes evaluation paths in the system that influenced the chosen assignment to have its value. We argue that showing such paths and not only the assignments of an inclusion-minimal cause provides useful visual information to the analyst.

\section{Case study}
\label{sec:case_study}
The main feature of \textsc{Oeritte} is a diagram view where it is possible to visually navigate through the causes of the assignment of interest to find out the roots of the system malfunction. 
That is why the tool should be evaluated by the analyst interacting with it to discover a problem in the real system.
As the latter, we use the mode selection logic introduced in~\cite{pakonen2017npic}. 
It contains an actual design issue, revealed by using model checking in a practical nuclear industry project.

\begin{figure*}[h]
\centerline{\includegraphics[width=0.9\textwidth]{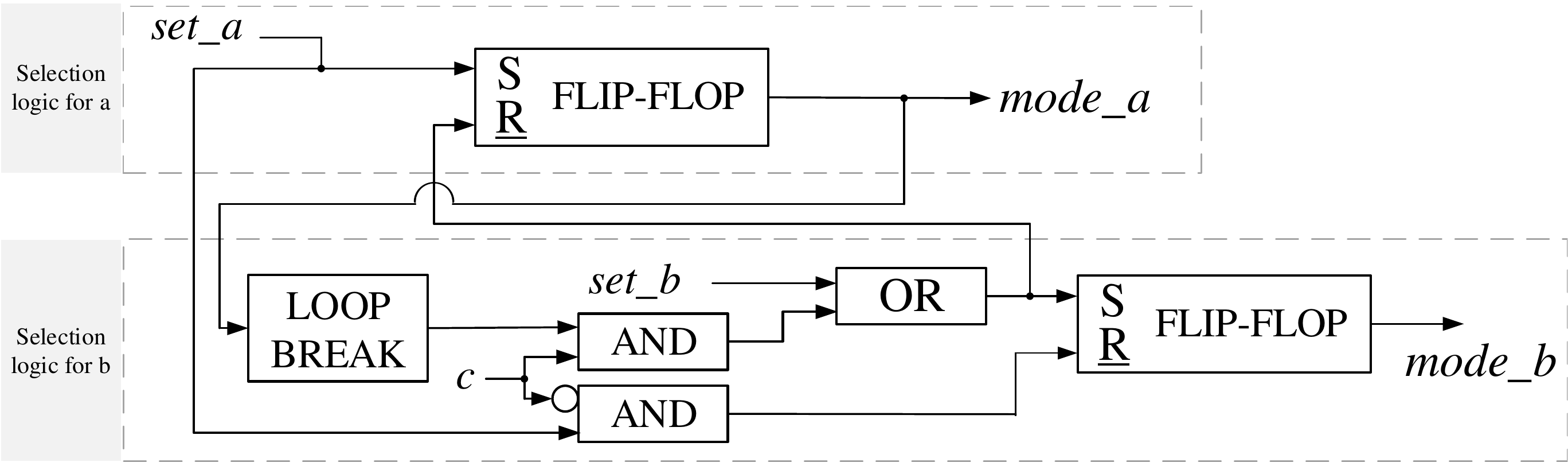}}
\caption{Case study: mode selection logic~\cite{pakonen2017npic}. \textit{Flip-flop} blocks produce \texttt{TRUE} if $\texttt{S} \land \neg \texttt{R}$ and \texttt{FALSE} otherwise. If both \texttt{S} and \texttt{R} are \texttt{FALSE} then the value from the previous time step is used as output. A \textit{loop breaker} works as a signal delay so that its output at the step \texttt{N} equals to its input at the step \texttt{N-1} and the predefined value for the step \texttt{0}, which is \texttt{FALSE} in our case. A \textit{circle} means that a block consumes the inverted input.
}
\label{fig:case_study}
\end{figure*}

The logic is schematically represented in Fig.~\ref{fig:case_study} and is used to select one of two operational modes, \texttt{mode\_a} or \texttt{mode\_b}. 
The operator can select the mode using the \texttt{set\_a} or \texttt{set\_b} command. 
In addition, if \texttt{mode\_a} is active, and the signal \texttt{c} then changes to true, the mode is automatically switched to \texttt{mode\_b}. 
The processing order for the feedback loop is specified using a cycle delay block.

One functional requirement for the logic is that the two modes shall not both be active at the same time. 
To accommodate for the cycle delay, the analyst specifies the LTL property $\mathbf{G}\neg((\mathtt{mode\_a} \land \mathtt{mode\_b}) \land \mathbf{X} (\mathtt{mode\_a} \land \mathtt{mode\_b}))$, i.e., \texttt{mode\_a} and \texttt{mode\_b} shall not both be on for two consecutive cycles. 
This specification is violated and NuSMV generates a counterexample.
The result of the explanation of the LTL formula is that both \texttt{mode\_a} and \texttt{mode\_b} are \texttt{TRUE} at the steps 2 and 3.

\begin{figure*}[tp]
\centering
    \subfloat[Explanation for the \texttt{mode\_b} output for the third step. To save space, in this figure we combined two screenshots with different tooltips being shown.] {
    	\centering
        \includegraphics[width=\textwidth,trim=0.2cm 0 0.3cm 0,clip]{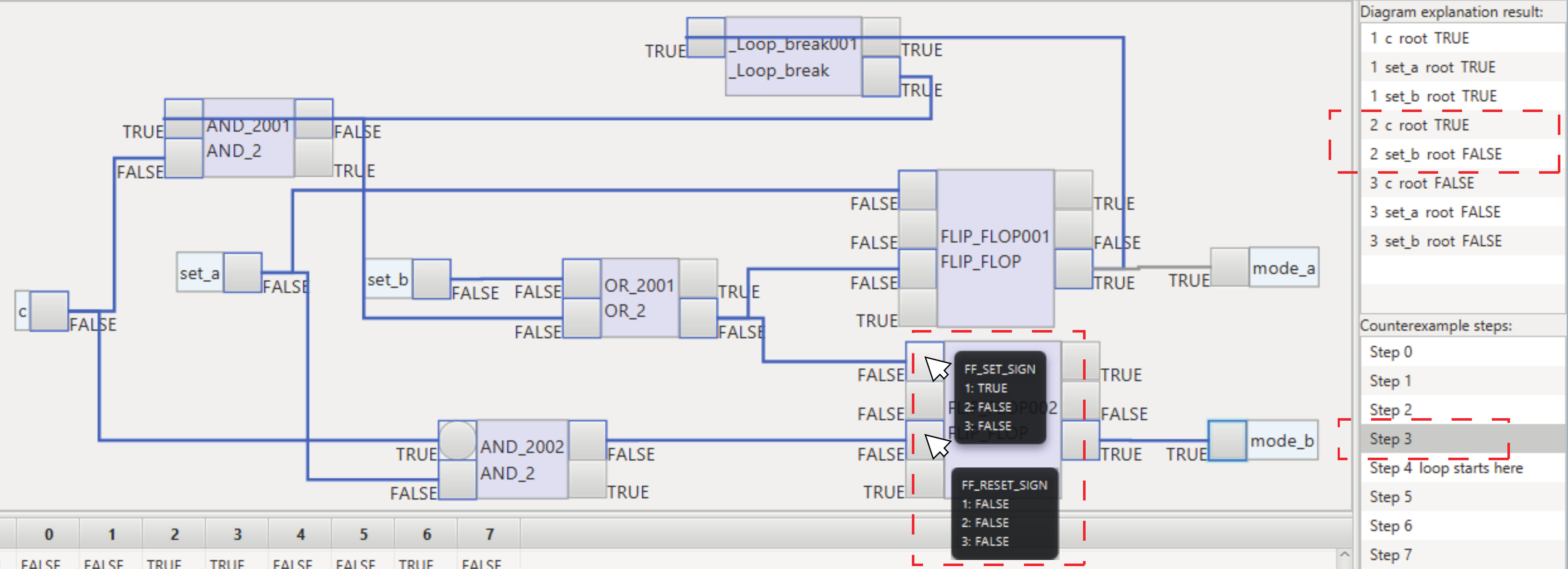}
        \label{fig:case_sequence:both_tooltips}
    }
	\qquad \qquad
    \subfloat[Explanation for the reset input of the second flip-flop block for the \emph{second} step. \texttt{c} is shown to be the cause of this signal not activated.] {
    	\centering 
        \includegraphics[width=\textwidth,trim=0.2cm 0 0.3cm 0,clip]{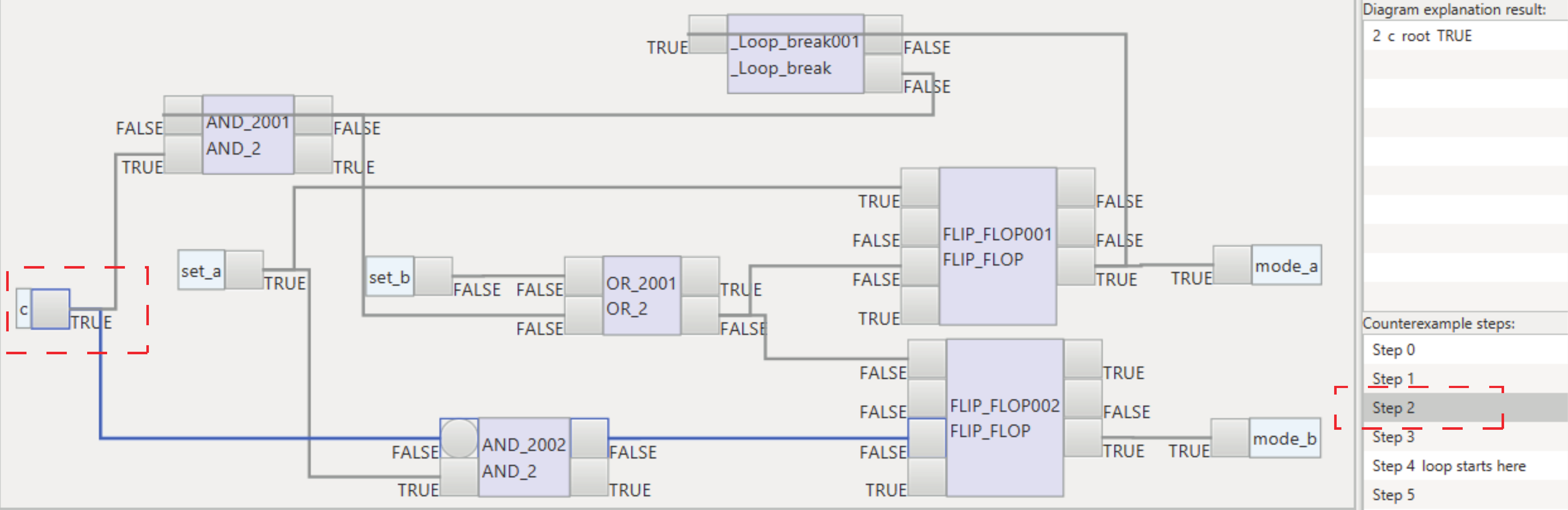}
        \label{fig:case_sequence:reset_2}
	}
	\qquad \qquad
	\subfloat[Explanation for the reset input of the second flip-flop block for the \emph{third} step. \texttt{set\_a} is shown to be the cause of this signal not activated.] {
    	\centering 
        \includegraphics[width=\textwidth,trim=0.2cm 0 0.3cm 0,clip]{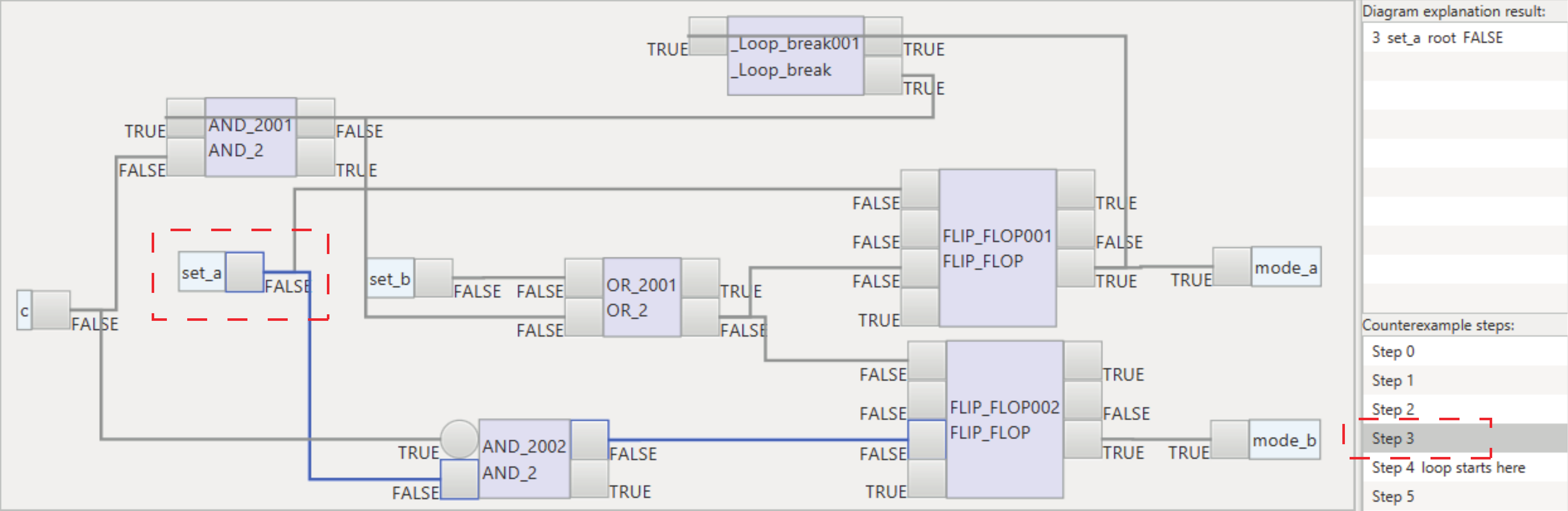}
        \label{fig:case_sequence:reset_3}
	}
    \caption{Counterexample explanation for the case study logic. The points of the analyst's attention are highlighted with red dashed rectangles and are not displayed by the tool.}
\end{figure*}

As this explanation does not contain information about the reasons for such system behavior, the analyst considers the diagram.
Having both modes active on the time steps~2 and~3, the analyst clicks on the \texttt{mode\_b} output at the time step~3, and the explanation for that value is shown.
Almost all diagram connections are highlighted, but, by hovering the mouse pointer over the pins, the analyst notices that the second flip-flop was set once at step 1 and never reset~(Fig.~\ref{fig:case_sequence:both_tooltips}).
The analyst then notices in the ``diagram explanation result'' panel that the \texttt{set\_a} command on time step 2 does \textit{not} have an effect, when, in fact, it should reset the \texttt{mode\_b} flip-flop a cycle later.
Then, the question is why the second flip-flop did not reset its output signal at the steps~2 and~3.
Clicks on its reset input at steps~2 and~3 outlines the path to \texttt{c}~(Fig.~\ref{fig:case_sequence:reset_2}) and to \texttt{set\_a}~(Fig.~\ref{fig:case_sequence:reset_3}) correspondingly, therefore, the analyst is shown that signal \texttt{c} is the cause at time step~2, and inactive \texttt{set\_a} command is the cause at step~3.

Combined, these results help the analyst realize that the issue is about an extremely short \texttt{set\_a} signal pulse arriving at a very specific time~-- one cycle before the signal \texttt{c} resets. 
(As explained in~\cite{pakonen2017npic}, the issue may also be counter-intuitive to the designer because invoking \texttt{set\_a} at a time when \texttt{c} is active is not necessarily feasible.)


\section{Related research}
\label{sec:rel}

There exist several approaches to counterexample explanation. Some works~\cite{ek2016explanation,beer2012,pakonen2018} focus on the failed property itself and, among mentioned,~\cite{pakonen2018} is the most advanced one, which was directly implemented in our work in the LTL formula explanation part. It generalizes an algorithm from~\cite{beer2012} that explains a false outcome of LTL formula $f$ on a given counterexample.
This is done by recursively finding assignments that are sufficient to cause the value of a subformula of $f$ on a given counterexample step, starting from the entire $f$ on the first step.


However, the problem of the research direction of these approaches is that it is only possible to get an explanation with the variables that are actually included in the LTL formula.
Unfortunately, these values may be influenced by others and the real cause of the violation of $f$ may be still hidden (as it was shown in Section~\ref{sec:case_study}), requiring the analyst to manually examine the formal model.

Direct analysis of a system under verification is considered in~\cite{Groce2003,groce2004understanding,leue2012counterexample,leitner2013causality}. These approaches require more than one system trace and, therefore, additional executions of supporting programs to acquire necessary counterexamples or traces that correspond to successful model runs. By 
contrast, in~\cite{wang2006whodunit} a single failure trace is enough to reason about the causes of the error, although 
here a counterexample is represented by a sequence of executions of program statements and such sequence is unknown in our scenario. Compared to the works above, our approach needs only one counterexample from a model checker to explain the failure. Despite being often developed in form of modular diagrams, I\&C systems require a more sophisticated visual approach for localizing failures which is provided by \textsc{Oeritte}.

Works~\cite{Jee2010,Loer2006,Patil2015} are related to counterexample visualization.
In~\cite{Jee2010} counterexamples are visualized using timing diagrams, however,
``model view'' is arguably~\cite{Loer2006} the most usable representation. 
MODCHK~\cite{pakonen2017esrel} is a graphical front-end for NuSMV that visualizes the counterexamples by animating the function block diagram. For use in nuclear applications, MODCHK also supports vendor-specific, non-standard function block types. Similarly to MODCHK, in~\cite{Patil2015}, counterexamples for IEC~61499 block diagrams are shown using a simulation model of the controlled process. Another graphical tool is the Simulink Design Verifier~\cite{simulink}, which generates a simulation test case from the counterexample.
Being able to visualize, the aforementioned tools do not support explanation of specifications failures.

A viewpoint that considers a structural approach to counterexample explanation and that is the closest to ours is taken in~\cite{bochot2010}, where the considered problem is to explain an output of an observer (synchronous program connected to the system under verification), which is a single value.
However, here only a limited set of basic blocks is supported, the approach is applied only to STANCE models and lacks flexibility in explanation scope determination. Another difference is that definitions in~\cite{bochot2010} are formulated over the paths in the system, where all those starting in system inputs require to obtain an activation condition formulae before the explanation starts. As a result, sets of paths representing a cause of the observer output is calculated.

In our work, we synthesize the ideas of the works~\cite{beer2012,pakonen2018,bochot2010} in the following form: we borrow the idea of explaining an LTL formula with particular values of variables in it, but also let the user get explanations for each such particular values.
We solve the comprehensibility problem by presenting only one set of paths and letting the user visually navigate through the FBD and the counterexample to get more focused explanations. Therefore, despite steps towards more user-friendly counterexample visualization and explanation have already been made, \textsc{Oeritte} is the only tool that combines explanation techniques into a consistent infrastructure. 

\section{Conclusion}
\label{sec:conclusion}
In this paper, we have presented a novel counterexample explanation algorithm and an open-source tool, \textsc{Oeritte}, which implements it together with a known LTL formula explanation algorithm~\cite{pakonen2018} and offers graphical backward counterexample analysis.
Inspired by works~\cite{beer2012,bochot2010}, the tool provides methods and visual elements supporting explanations in terms of both the LTL formula and the model (FBD) in a form of paths from causes to the target values that they explain. It is worth mentioning that our approach can be used to explain the results of finite computations even when they are produced by models which are not FBDs. 
The elements of the user interface of \textsc{Oeritte} and their functionality address the challenges mentioned in the introduction. The variables of the diagram~(Fig.~\ref{fig:block:visual}b, Fig.~\ref{fig:block:visual}c) and the LTL formula tree (Fig.~\ref{fig:main_view}a) are evaluated according to a chosen counterexample step, which addresses the first challenge. The second challenge is covered by the possibility to retrieve causes of the failure using only the formula structure, where the LTL formula tree~(Fig.~\ref{fig:main_view}a), the button ``explain formula'' and highlighted values (Fig.~\ref{fig:main_view}c) help with visualization. Finally, the diagram (Fig.~\ref{fig:main_view}b) combined with the presented method of individual assignment explanation assists in the analysis of the system as a whole.


As shown by the case study, the current version of \textsc{Oeritte} already avails, moreover, it has potential in reaching a wider audience in the 
future.
Our future work includes broadening the scope of supported models and developing the counterexample explanation methods for timed and probabilistic automata.
Another enhancement that would enlarge the range of models and specifications to be analyzed is relaxing input data restrictions, i.e., introducing support for bounded and past time LTL operators as well as removing necessity to specify the data types in a NuSMV model.  


\section*{Acknowledgments}
This work was supported, in part, by the Finnish Research Programme on Nuclear Power Plant Safety 2018-2022 (SAFIR 2022) and by the Government of the Russian Federation under Grant 08-08.

\bibliographystyle{IEEEtran}
\bibliography{main}

\begin{thebibliography}{10}
\providecommand{\url}[1]{#1}
\csname url@samestyle\endcsname
\providecommand{\newblock}{\relax}
\providecommand{\bibinfo}[2]{#2}
\providecommand{\BIBentrySTDinterwordspacing}{\spaceskip=0pt\relax}
\providecommand{\BIBentryALTinterwordstretchfactor}{4}
\providecommand{\BIBentryALTinterwordspacing}{\spaceskip=\fontdimen2\font plus
\BIBentryALTinterwordstretchfactor\fontdimen3\font minus
  \fontdimen4\font\relax}
\providecommand{\BIBforeignlanguage}[2]{{%
\expandafter\ifx\csname l@#1\endcsname\relax
\typeout{** WARNING: IEEEtran.bst: No hyphenation pattern has been}%
\typeout{** loaded for the language `#1'. Using the pattern for}%
\typeout{** the default language instead.}%
\else
\language=\csname l@#1\endcsname
\fi
#2}}
\providecommand{\BIBdecl}{\relax}
\BIBdecl

\bibitem{clarke1999}
E.~M. Clarke, O.~Grumberg, and D.~Peled, \emph{Model checking}.\hskip 1em plus
  0.5em minus 0.4em\relax MIT press, Cambridge, Massachusetts, 1999.

\bibitem{kaivola2009intel}
R.~Kaivola, R.~Ghughal, N.~Narasimhan, A.~Telfer, J.~Whittemore, S.~Pandav,
  A.~Slobodov{\'a}, C.~Taylor, V.~Frolov, E.~Reeber \emph{et~al.}, ``Replacing
  testing with formal verification in {Intel} \textsuperscript{\textregistered}
  {CoreTM} i7 processor execution engine validation,'' in \emph{International
  Conference on Computer Aided Verification}.\hskip 1em plus 0.5em minus
  0.4em\relax Springer, 2009, pp. 414--429.

\bibitem{nakahori2017iot}
K.~Nakahori and S.~Yamaguchi, ``A support tool to design {IoT} services with
  {NuSMV},'' in \emph{2017 IEEE International Conference on Consumer
  Electronics (ICCE)}.\hskip 1em plus 0.5em minus 0.4em\relax IEEE, 2017, pp.
  80--83.

\bibitem{pakonen2017npic}
A.~Pakonen, T.~Tahvonen, M.~Hartikainen, and M.~Pihlanko, ``Practical
  applications of model checking in the {Finnish} nuclear industry,'' in
  \emph{10th International Topical Meeting on Nuclear Plant Instrumentation,
  Control and Human Machine Interface Technologies (NPIC \& HMIT)}.\hskip 1em
  plus 0.5em minus 0.4em\relax American Nuclear Society, 2017, pp. 1342--1352.

\bibitem{Jee2010}
E.~Jee, S.~Jeon, S.~Cha, K.~Koh, J.~Yoo, G.~Park, P.~Seong \emph{et~al.},
  ``{FBDVerifier}: Interactive and visual analysis of counter-example in formal
  verification of function block diagram,'' \emph{Journal of Research and
  Practice in Information Technology}, vol.~42, no.~3, p. 171, 2010.

\bibitem{nemeth2009}
E.~N{\'e}meth and T.~Bartha, ``Formal verification of safety functions by
  reinterpretation of functional block based specifications,'' in
  \emph{International Workshop on Formal Methods for Industrial Critical
  Systems (FMICS 2008)}.\hskip 1em plus 0.5em minus 0.4em\relax Springer, 2008,
  pp. 199--214.

\bibitem{adiego2015}
B.~F. Adiego, D.~Darvas, E.~B. Vi{\~n}uela, J.-C. Tournier, S.~Bliudze, J.~O.
  Blech, and V.~M.~G. Su{\'a}rez, ``Applying model checking to industrial-sized
  {PLC} programs,'' \emph{IEEE Transactions on Industrial Informatics},
  vol.~11, no.~6, pp. 1400--1410, 2015.

\bibitem{gelmanAirbus2013}
G.~E. Gelman, K.~M. Feigh, and J.~Rushby, ``Example of a complementary use of
  model checking and agent-based simulation,'' in \emph{2013 IEEE International
  Conference on Systems, Man, and Cybernetics}.\hskip 1em plus 0.5em minus
  0.4em\relax IEEE, 2013, pp. 900--905.

\bibitem{IEC61131}
{International Electrotechnical Commission}, \emph{International Standard {IEC}
  61131-3:2013: Programmable Controllers. Part 3: Programming Languages}.\hskip
  1em plus 0.5em minus 0.4em\relax IEC, 2013.

\bibitem{lee2016introduction}
E.~A. Lee and S.~A. Seshia, \emph{Introduction to embedded systems: A
  cyber-physical systems approach}.\hskip 1em plus 0.5em minus 0.4em\relax MIT
  Press, 2016.

\bibitem{pakonen2017esrel}
A.~Pakonen and K.~Bj{\"o}rkman, ``Model checking as a protective method against
  spurious actuation of industrial control systems,'' in \emph{27th European
  Safety and Reliability Conference (ESREL 2017)}.\hskip 1em plus 0.5em minus
  0.4em\relax Taylor \& Francis Group, London, UK, 2017, pp. 3189--3196.

\bibitem{simulink}
``Simulink design verifier,''
  \url{https://www.mathworks.com/products/simulink-design-verifier.html}, last
  accessed 16 Sep 2020.

\bibitem{nusmv}
A.~Cimatti, E.~Clarke, E.~Giunchiglia, F.~Giunchiglia, M.~Pistore, M.~Roveri,
  R.~Sebastiani, and A.~Tacchella, ``{NuSMV} 2: An {OpenSource} tool for
  symbolic model checking,'' in \emph{International Conference on Computer
  Aided Verification (CAV)}.\hskip 1em plus 0.5em minus 0.4em\relax Springer,
  2002, pp. 359--364.

\bibitem{jee2010fbdverifier}
E.~Jee, S.~Jeon, S.~Cha, K.~Koh, J.~Yoo, G.~Park, P.~Seong \emph{et~al.},
  ``{FBDVerifier}: {Interactive} and visual analysis of counter-example in
  formal verification of function block diagram,'' \emph{Journal of Research
  and Practice in Information Technology}, vol.~42, no.~3, p. 171, 2010.

\bibitem{pakonen2018}
A.~Pakonen, I.~Buzhinsky, and V.~Vyatkin, ``Counterexample visualization and
  explanation for function block diagrams,'' in \emph{2018 IEEE 16th
  International Conference on Industrial Informatics (INDIN)}.\hskip 1em plus
  0.5em minus 0.4em\relax IEEE, 2018, pp. 747--753.

\bibitem{beer2012}
I.~Beer, S.~Ben-David, H.~Chockler, A.~Orni, and R.~Trefler, ``Explaining
  counterexamples using causality,'' \emph{Formal Methods in System Design},
  vol.~40, no.~1, pp. 20--40, 2012.

\bibitem{halpern2015modification}
J.~Y. Halpern, ``A modification of the {H}alpern-{P}earl definition of
  causality,'' \emph{arXiv preprint arXiv:1505.00162}, 2015.

\bibitem{ek2016explanation}
A.~Ek, ``Explanation of counterexamples in the context of formal verification,
  {B.S. Thesis, Uppsala University, Department of Information Technology},''
  2016.

\bibitem{Groce2003}
A.~Groce and W.~Visser, ``What went wrong: Explaining counterexamples,'' in
  \emph{International {SPIN} Workshop on Model Checking of Software}.\hskip 1em
  plus 0.5em minus 0.4em\relax Springer, 2003, pp. 121--136.

\bibitem{groce2004understanding}
A.~Groce, D.~Kroening, and F.~Lerda, ``Understanding counterexamples with
  explain,'' in \emph{International Conference on Computer Aided
  Verification}.\hskip 1em plus 0.5em minus 0.4em\relax Springer, 2004, pp.
  453--456.

\bibitem{leue2012counterexample}
S.~Leue and M.~T. Befrouei, ``Counterexample explanation by anomaly
  detection,'' in \emph{International SPIN Workshop on Model Checking of
  Software}.\hskip 1em plus 0.5em minus 0.4em\relax Springer, 2012, pp. 24--42.

\bibitem{leitner2013causality}
F.~Leitner-Fischer and S.~Leue, ``Causality checking for complex system
  models,'' in \emph{International Workshop on Verification, Model Checking,
  and Abstract Interpretation}.\hskip 1em plus 0.5em minus 0.4em\relax
  Springer, 2013, pp. 248--267.

\bibitem{wang2006whodunit}
C.~Wang, Z.~Yang, F.~Ivan{\v{c}}i{\'c}, and A.~Gupta, ``Whodunit? causal
  analysis for counterexamples,'' in \emph{International Symposium on Automated
  Technology for Verification and Analysis}.\hskip 1em plus 0.5em minus
  0.4em\relax Springer, 2006, pp. 82--95.

\bibitem{Loer2006}
K.~Loer and M.~D. Harrison, ``An integrated framework for the analysis of
  dependable interactive systems ({IFADIS}): Its tool support and evaluation,''
  \emph{Automated Software Engineering}, vol.~13, no.~4, pp. 469--496, 2006.

\bibitem{Patil2015}
S.~Patil, V.~Vyatkin, and C.~Pang, ``Counterexample-guided simulation framework
  for formal verification of flexible automation systems,'' in \emph{2015 IEEE
  13th International Conference on Industrial Informatics (INDIN)}.\hskip 1em
  plus 0.5em minus 0.4em\relax IEEE, 2015, pp. 1192--1197.

\bibitem{bochot2010}
T.~Bochot, P.~Virelizier, H.~Waeselynck, and V.~Wiels, ``Paths to property
  violation: A structural approach for analyzing counter-examples,'' in
  \emph{2010 IEEE 12th International Symposium on High Assurance Systems
  Engineering}.\hskip 1em plus 0.5em minus 0.4em\relax IEEE, 2010, pp. 74--83.

\end{thebibliography}

\end{document}